\def\year{2025}\relax
\documentclass[letterpaper]{article} 
\usepackage{aaai25}  
\usepackage{times}  
\usepackage{helvet}  
\usepackage{courier}  
\usepackage[hyphens]{url}  
\usepackage{graphicx} 
\usepackage{natbib}  
\usepackage{caption} 
\usepackage{algorithm}
\usepackage{algorithmic}

\usepackage{booktabs}
\usepackage{xcolor}
\usepackage{amsmath}
\usepackage{bm}
\usepackage{amsfonts}
\usepackage{multicol}
\usepackage{graphicx}
\usepackage{multirow}

\usepackage{newfloat}
\usepackage{listings}
\DeclareCaptionStyle{ruled}{labelfont=normalfont,labelsep=colon,strut=off} 
\floatstyle{ruled}
\newfloat{listing}{tb}{lst}{}
\floatname{listing}{Listing}
\title{Gesture-Aware Zero-Shot Speech Recognition\\for Patients with Language Disorders}

\author {
    Seungbae Kim\textsuperscript{\rm 1}\footnote{Equal contribution.},
    Daeun Lee\textsuperscript{\rm 2}\footnotemark[1],
    Brielle Stark\textsuperscript{\rm 3},
    Jinyoung Han\textsuperscript{\rm 4}\footnote{Corresponding author.}
}
\affiliations {
    \textsuperscript{\rm 1} Department of Computer Science and Engineering, University of South Florida\\
    \textsuperscript{\rm 2} Yale School of Medicine, Yale University\\
    \textsuperscript{\rm 3} Department of Speech, Language and Hearing Sciences, Indiana University Bloomington\\
    \textsuperscript{\rm 4} Department of Applied Artificial Intelligence, Sungkyunkwan University\\
    seungbae@usf.edu, da-eun.lee@yale.edu, bcstark@iu.edu, jinyounghan@skku.edu
}

\begin{document}

\maketitle

\begin{abstract} 
Individuals with language disorders often face significant communication challenges due to their limited language processing and comprehension abilities, which also affect their interactions with voice-assisted systems that mostly rely on Automatic Speech Recognition (ASR). Despite advancements in ASR that address disfluencies, there has been little attention on integrating non-verbal communication methods, such as gestures, which individuals with language disorders substantially rely on to supplement their communication. Recognizing the need to interpret the latent meanings of visual information not captured by speech alone, we propose a gesture-aware ASR system utilizing a multimodal large language model with zero-shot learning for individuals with speech impairments. Our experiment results and analyses show that including gesture information significantly enhances semantic understanding. This study can help develop effective communication technologies, specifically designed to meet the unique needs of individuals with language impairments.
\end{abstract}

\section{Introduction}

Language disorders, such as aphasia, arise from damage to brain regions responsible for language production and comprehension. Aphasia is most often caused by acquired brain injuries, like stroke, and persists chronically in at least 30\%\ of cases~\cite{broca1861remarks,wasay2014stroke}. This means that many are living with aphasia -- indeed, there are nearly two million with aphasia in the USA alone~\cite{simmons2018aphasia}. Individuals with language disorders face significant communication challenges due to difficulties in processing and understanding language, resulting in impaired social interactions and reduced quality of life~\cite{el2017screening}. As voice-assisted technologies like Siri and Alexa become integral to daily activities, the inability of individuals with language disorders to interact effectively with these systems exacerbates their communication barriers, leading to frustration and further marginalization~\cite{rohlfing2021hey}.

Automatic Speech Recognition (ASR) systems are designed to transcribe spoken language into text, serving as a cornerstone for many voice-driven technologies. While recent advancements have made ASR systems more adept at handling disfluencies such as stuttering, their performance heavily depends on clear and consistent audio input~\cite{radford2023robust}. For individuals with language disorders, speech distortions, substitutions, and disjointed delivery pose significant challenges to accurate ASR transcription~\cite{sanguedolce2023uncovering}. As a result, current ASR solutions often fall short in addressing the needs of this population.

To enhance the robustness of ASR, researchers have explored Audio-Visual Speech Recognition (AVSR) systems that combine auditory and visual information~\cite{cheng2023opensr}. These systems leverage visual features such as lip movement~\cite{hu2023hearing,cheng2023mixspeech} and facial expressions~\cite{zadeh2016multimodal,busso2008iemocap} to improve speech recognition accuracy. However, these approaches are not ideal for individuals with language disorders, who often experience concomitant motor speech disorders and facial hemiplegia, which can result in 'masked' facial expressions~\cite{multani2017emotion,duffy2012motor}.

In contrast, individuals with language disorders frequently rely on non-verbal communication, such as gestures, to compensate for their verbal limitations~\cite{stark2023demographic}. Iconic gestures, in particular, serve as powerful tools for conveying meaning when spoken language is insufficient~\cite{de2023does,van2017production,stark2022task,stark2023demographic}. Unlike speech or facial expressions, iconic gestures offer a visual representation of concepts, enabling individuals with language impairments to express ideas more effectively~\cite{lee2023learning}. However, current ASR and AVSR systems fail to consider the latent semantic information encoded in these gestures, leaving a critical gap in understanding the full context of communication for individuals with language disorders.

To address the limitations of current ASR and AVSR systems, we introduce a gesture-aware zero-shot speech recognition framework specifically designed for individuals with language disorders. Our approach harnesses the capabilities of multimodal large language models (LLMs) to incorporate linguistic, acoustic, and gestural information, enabling a deeper understanding of the speaker's intended meaning. Unlike traditional systems that rely solely on spoken or visual cues like lip movements, our method emphasizes the integration of iconic hand gestures—gestures that visually represent concepts—to enrich the transcription process. This allows the system to produce transcripts that not only reflect spoken content but also capture the latent meanings conveyed through gestures, which are crucial for individuals with impaired speech.

The core of our method lies in leveraging a zero-shot framework to align and synthesize multimodal inputs. Our system processes disfluent or incomplete speech signals while simultaneously analyzing visual data to identify and interpret gestures. By fusing these streams of information, the system generates semantically enriched transcripts that bridge the gaps left by speech alone. This zero-shot design eliminates the need for task-specific training, making the approach adaptable to a wide range of scenarios and linguistic contexts. This adaptability is particularly advantageous for language-disordered populations, where speech patterns and gestures vary widely between individuals.

Our experiments show that our proposed model successfully generates transcripts by incorporating significant gestural information to ascertain the latent intent of individuals with language disorders, which is not conveyed through speech alone. This findings indicate that incorporating non-verbal information can effectively aid in advancing the creation of more inclusive and effective communication technologies designed specifically for the distinctive requirements of individuals with language impairments.

\section{Related Work}

\subsection{ASR as Assistive Technologies}
Assistive technologies play a critical role in supporting clinicians to deliver assessments and therapies, as well as in facilitating communication for individuals with aphasia. In speech-language pathology, such tools are often referred to as augmentative and alternative communication (AAC) systems~\cite{beukelman1998augmentative}. ASR technology has shown promise in enhancing communication by enabling real-time, accurate feedback to individuals with aphasia~\cite{ballard2019feasibility,barbera2021nuva}. This is particularly valuable as individuals with aphasia frequently face challenges with self-monitoring~\cite{oomen2001prearticulatory,sampson2011investigation} and often benefit from external cues to improve their performance~\cite{tompkins2006communicative,conroy2009effects,schwartz2016does}. Additionally, ASR can improve communication efficiency and quality by compensating for writing or grammatical impairments through speech-to-text conversion and delivering feedback during therapy~\cite{ballard2019feasibility,barbera2021nuva}. Consequently, ASR-based applications have gained popularity in delivering speech-language services for individuals with aphasia and other neurogenic conditions, such as Parkinson’s disease~\cite{hoover2014integrating,mccrocklin2016pronunciation,strik2009comparing}.

However, despite advancements in ASR, reliance on voice-based systems alone often results in transcription errors when processing disfluent or incomplete speech, a hallmark of language impairments~\cite{jefferson2019usability,le2016automatic}. These inaccuracies can hinder communication and reduce the overall effectiveness of these tools, further exacerbating barriers faced by individuals with language disorders. To address these challenges, AVSR systems have emerged, combining visual inputs—such as lip movements and facial expressions—with audio to enhance recognition accuracy~\cite{gabeur2022avatar,afouras2018deep,dupont2000audio,ma2021end,noda2015audio,mroueh2015deep,feng2017audio}. However, AVSR systems often fall short for individuals with language and speech disorders. Speech disorders such as dysarthria and apraxia frequently disrupt articulation, rendering lip movements unreliable. Moreover, neutral or ambiguous facial expressions common in these populations provide limited contextual insight~\cite{tong2020automatic,salama2014audio}.

\begin{figure*}[ht]
    \centering
    \includegraphics[width=\linewidth]{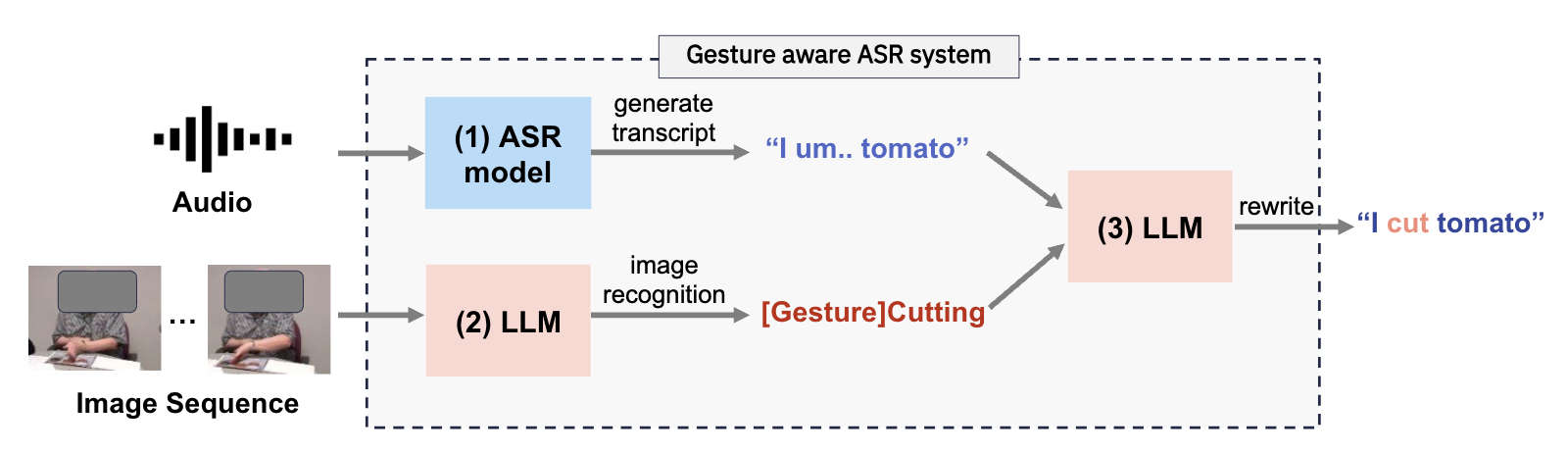}
    \caption{The overall process of the proposed system. Our model integrates incomplete speech and visual data (i.e., iconic gestures) and generates semantically enriched transcripts.}
    ~\label{fig:framework}
\end{figure*}

\subsection{Gesture into Assistive Technology}
Gestures play a vital role in communication, often conveying critical information that complements, enhances, or even substitutes spoken language~\cite{kita2009cross,kendon1994gestures,mcneill1990speech}. This holds true for individuals with and without language impairments, as gestures are co-produced with speech across languages and cultures, aiding both speakers and listeners~\cite{kita2009cross}. Remarkably, even congenitally blind speakers, who have never observed gestures, use gestures as frequently as sighted individuals, emphasizing their fundamental role in communication~\cite{iverson1998people}. For speakers, gestures illustrate abstract concepts, emphasize key points, and provide additional information that speech alone may struggle to convey~\cite{mcneill1990speech,kita2000representational,goldin2013gesture}. For listeners, gestures offer visual cues that complement verbal messages, enhancing understanding in noisy settings, during non-native language processing, and when speech is ambiguous~\cite{cook2009embodied,goldin1999role}. Notably, listeners often recall information received via gestures as if it were spoken, underscoring their semantic importance~\cite{cassell2000embodied,kelly1999offering}.

Among gestures, representational gestures—those that visually represent objects or actions—are particularly valuable for enhancing communication~\cite{mcneill1990speech}. Iconic gestures, a subtype of representational gestures, depict concrete referents (e.g., a kicking motion to signify “kick”)~\cite{novack2017gesture}. Other representational gestures include pantomimes, which occur without speech, and deictic gestures, such as pointing, which may emphasize or add meaning to speech\cite{kendon1994gestures}. Additionally, emblems like a “thumbs up” convey culturally specific meanings independent of speech~\cite{kendon1994gestures}. However, not all gestures are communicative. Non-communicative movements, such as tucking hair behind the ear, do not contribute to the conveyed message and can be irrelevant to the communicative process.

For assistive technologies, particularly AI-based systems, distinguishing communicative gestures from non-communicative movements is essential. Systems must accurately interpret gestures to capture the full intent of a speaker’s message, especially when assisting individuals with language impairments. Thus, effectively integrating linguistic and co-speech gesture information is critical for improving the utility and effectiveness of AI-driven assistive tools.

\subsection{Challenges in Gesture Interpretation for Assistive Technology}
Accurately interpreting the wide variety of gestures individuals use is a significant challenge, as gestures vary in form and meaning, with some being non-communicative. Even gestures conveying the same intent, such as cutting, can differ—for example, an “enacting” gesture mimics the action of cutting with a flat hand, while a “handling” gesture simulates gripping an invisible knife~\cite{poggi2008iconicity,hassemer2018decoding}. This variability complicates consistent interpretation, particularly in zero-shot settings. While recent computer vision methods focus on generating gestures from text inputs for avatars or video sequences~\cite{ginosar2019learning,ahuja2022low,liu2022beat}, they do not address gesture understanding or intent recognition, relying instead on predefined datasets. Existing ASR and AVSR systems also fall short, as they lack the ability to incorporate conversational context or personalized knowledge, both critical for interpreting gestures tied to specific topics or individual habits. For instance, one person may favor “handling” gestures while another prefers “enacting.” Additionally, task-specific training required by many systems is impractical for real-time applications, as it necessitates frequent retraining to accommodate new users and scenarios. These limitations underscore the need for adaptable systems capable of real-time, personalized gesture interpretation.

\section{Method}
Given that the dataset provides video recordings of persons with language disorders, the proposed model aims to improve the transcription of audio from individuals with language disorders, focusing on enhancing contextual understanding, incorporating gesture, and resolving ambiguities in speech. It consists of three interconnected components: Speech Recognition, Gesture Recognition, and Contextual Rewriting. Figure~\ref{fig:framework} provides a schematic overview of the system. 

\subsubsection{Speech Recognition}
The first step in the pipeline is audio processing using an ASR system. This component takes an audio signal $\mathbf{A}$ as input and outputs a preliminary transcript $T_{\text{ASR}}$, which represents the system’s initial interpretation of spoken words. Let the ASR model be represented by $f_{\text{ASR}}$ as follows,
\begin{equation}
T_{\text{ASR}} = f_{\text{ASR}}(\mathbf{A}), \quad T_{\text{ASR}} = \{ w_1, w_2, \ldots, w_n \}
\end{equation}
where $w_i$ is the $i$-th word in the transcript.
Consider an audio input from a patient saying, ``I um... tomato.'' The ASR model may generate a raw transcript as incomplete speech. This transcript lacks clarity and completeness, as it fails to convey the full meaning intended by the speaker. The limitations arise because ASR systems rely solely on the audio signal and are unable to incorporate accompanying non-verbal cues, such as gestures, that could provide additional context.

\subsubsection{Gesture Recognition}   
The second component focuses on identifying gestures from the video frames corresponding to the speech. The input is a sequence of video frames $\mathbf{V} = \{ v_1, v_2, \ldots, v_m \}$, where each frame $v_i$ captures a snapshot of the speaker's hand or body movements. A gesture recognition model, $f_{\text{Gesture}}$, processes these frames to detect meaningful gestures as follows,
\begin{equation}
\mathbf{G} = f_{\text{Gesture}}(\mathbf{V}), \quad \mathbf{G} = \{ g_1, g_2, \ldots, g_k \}
\end{equation}
where $g_i$ represents a detected gesture and its associated meaning. Due to the scarcity of labeled datasets for iconic gestures, the framework employs a multimodal large language model to perform zero-shot gesture recognition. This approach allows the model to infer the meanings of gestures based on generalizable patterns learned from other modalities. Iconic gestures, which represent concrete referents, are of particular interest because they often clarify or complement the spoken language.

If the video captures the speaker making a back-and-forth motion with a flat hand, the gesture recognition model identifies this as a cutting motion and assigns it the label $g$=``cut''. Iconic gestures, however, can vary significantly between individuals. For instance, one person might perform a ``handling'' gesture by mimicking the act of holding and moving an invisible knife, while another might use an ``enacting'' gesture, simulating the motion of cutting without mimicking the act of holding. The proposed model is designed to handle such variability by recognizing these different representations and mapping them to a unified semantic meaning, ensuring that the gesture's intent is accurately interpreted regardless of individual differences in gesture style.

\subsubsection{Contextual Rewriting}
The third component integrates the outputs from the ASR and gesture recognition models to generate a contextually enriched and semantically accurate transcript. This process is handled by a LLM which takes the initial transcript $T_{\text{ASR}}$, the detected gestures $\mathbf{G}$ as inputs. The final transcript, $T_{\text{Final}}$, is generated as follows:
\begin{equation}
T_{\text{Final}} = f_{\text{LLM}}(T_{\text{ASR}}, \mathbf{G}), \quad T_{\text{Final}} = \{ w'_1, w'_2, \ldots, w'_p \}
\end{equation}
where $w'_i$ represents a word in the final transcript. The model directly uses the recognized gestures as the sole non-verbal input, streamlining the process while focusing on the integration of multimodal signals.

For example, if the ASR transcript reads, "I um... tomato," and the gesture recognition model identifies the label $g$=``cut''
from a cutting motion, the LLM rewrites the transcript as ``I cut tomato.'' This correction ensures that the final output reflects both the spoken and gestural communication, improving the system's ability to understand and represent the speaker's intent accurately.
This approach enables the generation of accurate and enriched transcripts, even in cases where speech alone might be ambiguous or incomplete. By leveraging the co-speech gesture input, the system enhances its interpretive power without requiring additional external context.

\begin{table}[t]
\centering
\caption{The corpora from AphasiaBank}
\vspace{-0.1in}
\label{tab:appendix_corpus}
\resizebox{\linewidth}{!}{%
\begin{tabular}{c|c}
\hline
Corpus                                                                & Site                                       \\ \hline \hline
ACWT~\cite{bynek2013aphasiabank}                & Aphasia Center of West Texas               \\
Adler~\cite{adler2013aphasia}                   & Adler Aphasia Center                       \\
APROCSA~\cite{aprocsa2021aphasia}               & Vanderbilt University Medical Center       \\
BU~\cite{bu2013apha}                            & Boston University                          \\
Capilouto~\cite{capilouto2008apha}              & University of Kentucky                     \\
CC-Stark~\cite{cc2022apha}                      & file for CC                                \\
CMU~\cite{cmu2013apha}                          & Carnegie Mellon University                 \\
Elman~\cite{elman2011starting,elman2016aphasia} & Aphasia Center of California               \\
Fridriksson~\cite{Fridriksson2013apha}          & University of South Carolina               \\
Garrett~\cite{Garrett2013apha}                  & Pittsburgh, PA                             \\
Kansas~\cite{kansas2013apha}                    & University of Kansas                       \\
Kempler~\cite{kempler2013apha}                  & Emerson College                            \\
Kurland~\cite{kurland2013apha}                  & University of Massachusetts, Amherst       \\
MSU~\cite{MSU2013apha}                          & Montclair State University                 \\
NEURAL~\cite{neural2023apha}                    & NEURAL Research Lab, Indiana University    \\
Richardson~\cite{Richardson2008apha}            & University of New Mexico                   \\
SCALE~\cite{scale2013apha}                      & Snyder Center for Aphasia Life Enhancement \\
STAR~\cite{star2013apha}                        & Stroke Aphasia Recovery Program            \\
TAP~\cite{tap2013apha}                          & Triangle Aphasia Project                   \\
TCU~\cite{tcu2013apha}                          & Texas Christian University                 \\
Thompson~\cite{thompson2013apha}                & Northwestern University                    \\
Tucson~\cite{tucson2013apha}                    & University of Arizona                      \\
UCL~\cite{ucl2021apha}                          & University College London                  \\
UMD~\cite{faroqi2018comparison}                 & University of Maryland                     \\
UNH~\cite{unh2013apha}                          & University of New Hampshire                \\
Whiteside~\cite{whiteside2013apha}              & University of Central Florida              \\
Williamson~\cite{williamson2013apha}            & Stroke Comeback Center                     \\
Wozniak~\cite{wozniak2013apha}                  & InteRACT                                   \\
Wright~\cite{wright2013apha}                    & Arizona State University                   \\ \hline
\end{tabular}%
}
\end{table}

\section{Experiments}
We collected the dataset from AphasiaBank~\cite{macwhinney2011aphasiabank, forbes2012aphasiabank}, a shared database created by clinical experts for aphasia research; the corpus information is summarized in Table~\ref{tab:appendix_corpus}. The dataset includes video recordings of the language evaluation test process between a pathologist and a subject, which also contains human-annotated transcriptions and subjects' demographic information.

\begin{table}[t]
\centering
\caption{Summary of data statistics of `Peanut Butter Task' based on gesture annotation.}
\label{tab:data_statistics}
\resizebox{\linewidth}{!}{%
\begin{tabular}{@{}cccccc@{}}
\toprule
\multirow{2}{*}{label}  & \multirow{2}{*}{\# utt} & \multirow{2}{*}{\# user} & \multicolumn{3}{c}{duration (sec)} \\ \cmidrule(l){4-6} 
                        &                         &                          & mean      & min       & max        \\ \midrule
{[}gesture:cutting{]}   & 109                     & 92                       & 3.76      & 0.21      & 18.17      \\
{[}gesture:eating{]}    & 54                      & 48                       & 2.01      & 0.12      & 20.79      \\
{[}gesture:folding{]}   & 159                     & 145                      & 2.95      & 0.09      & 21.31      \\
{[}gesture:layering{]}  & 45                      & 31                       & 3.92      & 0.39      & 16.99      \\
{[}gesture::opening{]}  & 34                      & 24                       & 4.56      & 0.31      & 16.29      \\
{[}gesture:spreading{]} & 269                     & 169                      & 4.53      & 0.14      & 26.42      \\ \midrule
total                   & 670                     & 288                      & 3.99      & 0.26      & 19.77      \\ \bottomrule
\end{tabular}%
}
\end{table}

For our study, we selected the ``Peanut Butter Sandwich Task'' as the focal activity where people are asked to explain the procedure for making a sandwich. This task has been demonstrated to associate with high rates of iconic gesturing~\cite{stark2022task,stark2023demographic,pritchard2015language,illes1989neurolinguistic}.

We performed a detailed analysis of the dataset to examine the distribution and characteristics of gestures used by participants during this task. Table~\ref{tab:data_statistics} provides a summary of the data, showing six distinct gesture types: cutting, spreading, folding, eating, layering, and opening. Each gesture corresponds to an action performed to describe or demonstrate a step in the sandwich-making process. For example, the cutting gesture involves a back-and-forth hand motion mimicking the act of slicing, while the spreading gesture typically represents the act of spreading condiments with a sweeping motion of the hand. The data highlights the natural integration of gestures with speech, making it a valuable resource for gesture-aware ASR systems.

Among the gestures, spreading was the most frequently used, appearing in 269 utterances across 169 users, with an average duration of 4.53 seconds. This gesture reflects a critical step in the sandwich-making process, as participants commonly describe spreading peanut butter or other ingredients. Following this, folding emerged as the second most frequently observed gesture, with 159 instances across 145 users, averaging 2.95 seconds in duration. The cutting gesture ranked third, appearing 109 times across 92 users, with an average duration of 3.76 seconds. These three gestures collectively account for a significant portion of the dataset, indicating their central role in describing sandwich preparation. These findings align with linguistic analysis of the same story, demonstrating cutting, spreading, and folding as core parts of the story~\cite{dalton2020compendium}.

The dataset also revealed interesting variability in gesture execution. For instance, while spreading gestures were consistently observed, their duration ranged from brief motions lasting just 0.14 seconds to extended actions of up to 26.42 seconds. Similarly, cutting gestures varied in duration, from 0.21 seconds to 18.17 seconds, highlighting differences in participants’ expressive styles and cognitive processing abilities. Gestures such as layering and opening were less frequently observed, occurring 45 and 34 times respectively, but exhibited average durations comparable to the more common gestures, suggesting that even less frequent gestures are articulated with similar complexity.

This analysis highlights the diversity in both the frequency and execution of gestures among participants, emphasizing the importance of robust recognition systems capable of capturing these variations. The prevalence of gestures like spreading, folding, and cutting also demonstrates their importance as key features for understanding communication in tasks involving procedural explanations. These findings provide a strong foundation for training and evaluating our gesture-aware ASR framework, which seeks to seamlessly integrate gesture information into speech transcription.

\section{Results}

We conducted experiments to evaluate the performance of ASR models in generating initial transcripts. These experiments are essential for assessing the quality of the ASR output, which serves as the input for our gesture-aware contextual rewriting model. Additionally, we performed a case study on selected samples to highlight the improvements made by our proposed approach compared to the initial ASR outputs.

\subsection{Speech Recognition Models}

\begin{table}[t]
\centering
\caption{The Average Word Error Rate (WER) between original transcript and ASR results.}\label{tab:wer}
\resizebox{0.7\linewidth}{!}{%
\begin{tabular}{@{}cc@{}}
\toprule
\textbf{Model}                                                         & WER                       \\ \midrule
\textbf{Whisper~\cite{radford2023robust}}        & 0.557                     \\
\textbf{Seamless~\cite{barrault2023seamlessm4t}} & 0.891                     \\
\textbf{Wave2Vec~\cite{baevski2020wav2vec}}      & 0.630                     \\
\textbf{Whisper (conf $>$ 0.2) }                   & 0.519 \\ \bottomrule
\end{tabular}%
}
\end{table}

The foundation of a successful gesture-aware ASR system lies in obtaining high-quality initial transcripts from the speech recognition stage. To this end, we compared state-of-the-art ASR models on the AphasiaBank dataset to determine their performance in generating accurate transcripts. The results, summarized in Table~\ref{tab:wer}, highlight the average Word Error Rate (WER) for each model, calculated by comparing their output with the ground truth human-annotated transcripts. WER serves as a critical metric for evaluating ASR systems, representing the proportion of errors (insertions, deletions, and substitutions) in the generated transcript relative to the total number of words in the ground truth.

Among the models tested, Whisper~\cite{radford2023robust} demonstrated the best performance, achieving the lowest WER of 0.557. This indicates that Whisper is particularly effective in handling speech from participants with aphasia, despite challenges such as disfluencies, atypical speech patterns, and background noise. Wave2Vec~\cite{baevski2020wav2vec} performed moderately well with a WER of 0.630, while Seamless~\cite{barrault2023seamlessm4t}, designed for multilingual ASR, showed a higher WER of 0.891, indicating potential limitations in its robustness to the specific speech characteristics in this dataset.

To further improve the accuracy of the initial transcripts, we applied a confidence-based filtering approach to Whisper's outputs. By utilizing Whisper’s token-level confidence scores, we excluded tokens with a confidence score below 0.2, thereby retaining only the most reliable portions of the transcript. This refinement reduced Whisper's WER from 0.557 to 0.519, further solidifying its position as the most accurate ASR model in our evaluations.

These results highlight the importance of selecting a robust ASR model for datasets featuring non-standard speech patterns. Whisper, particularly with confidence-based filtering, provides a strong baseline for generating initial transcripts. The reduced WER achieved by Whisper demonstrates its ability to capture key elements of speech even under challenging conditions, thereby enhancing the reliability of the downstream contextual rewriting process.

\subsection{A Case Study}

\begin{table}[t]
\centering
\caption{The result of a case study for comparison between Whisper and the proposed model. \textit{[gesture:]} indicates the type of iconic gestures used by speakers while speaking.}
\label{tab:case}
\resizebox{\linewidth}{!}{%
\begin{tabular}{@{}clll@{}}
\toprule
\textbf{Index} & \multicolumn{1}{c}{\textbf{Origianl}}                                                                    & \multicolumn{1}{c}{\textbf{Whisper}}                                   & \multicolumn{1}{c}{\textbf{Ours}}                                                               \\ \midrule
1              & \begin{tabular}[c]{@{}l@{}}w {[}gesture:layering{]} \\ is right.\end{tabular}                            & There's right there.                                                   & \begin{tabular}[c]{@{}l@{}}You put things right there, \\ doing a \textbf{layering} action.\end{tabular} \\
2              & \begin{tabular}[c]{@{}l@{}}{[}gesture:folding{]} uz@u \\ uh right yes.\end{tabular}                      & right.                                                                 & \textbf{folding} it right.                                                                               \\
3              & \begin{tabular}[c]{@{}l@{}}um {[}gesture:cutting{]} \\ {[}gesture:cutting{]} banana.\end{tabular}        & Um... Banana.                                                          & \textbf{cutting} banana.                                                                                 \\
4              & and {[}gesture:eating{]}.                                                                                & and                                                                    & and \textbf{eating}                                                                                      \\
5              & \begin{tabular}[c]{@{}l@{}}but a w knife \\ {[}gesture:spreading{]} \\ do the stuff like um\end{tabular} & \begin{tabular}[c]{@{}l@{}}night do this stuff \\ like oh\end{tabular} & \begin{tabular}[c]{@{}l@{}}do this night stuff \\ like oh \textbf{spreading}\end{tabular}                \\ \bottomrule
\end{tabular}%
}
\end{table}

To evaluate the effectiveness of the proposed multimodal model in capturing and incorporating the latent meaning conveyed by iconic gestures, we conducted a case study comparing its outputs with those of the Whisper ASR model. Table~\ref{tab:case} presents the results, showcasing the original transcripts in the dataset, the transcript generated by Whisper, and the transcript generated by our multimodal approach. The goal of this comparison was to determine whether our model can enhance transcript quality by inferring and integrating the semantic meaning of gestures observed during speech.

As shown in the table, Whisper provides a literal transcription of the spoken words but fails to capture the contextual or semantic information conveyed by the speaker's gestures. For example, in the first example, the original input includes the layering gesture, indicating an action of stacking or placing items on top of each other. Whisper transcribes this as ``\textit{There's right there}," which misses the gesture's implicit meaning. In contrast, our proposed model generates the enriched transcript, ``\textit{You put things right there, doing a layering action},'' effectively incorporating the gesture's semantic context into the spoken narrative.

In the second example, the speaker performs a folding motion, which is entirely absent from Whisper's transcription, resulting in the simple word ``\textit{right}." However, our model identifies and integrates this gesture into the transcript, producing ``\textit{folding it right}," which provides a more comprehensive description of the speaker's intention.

Similarly, in the third and fourth examples, iconic gestures like cutting and eating are completely omitted from Whisper's transcription, leading to outputs that lack essential contextual details. The proposed model enriches these outputs by explicitly including the actions, resulting in transcripts like ``\textit{cutting banana}" and ``\textit{and eating}," which more accurately reflect the speaker's message.

The fifth example highlights a limitation in handling more complex sentences where gestures provide significant contextual cues. While our model attempts to incorporate the spreading gesture into the generated transcript, the result, ``\textit{do this night stuff like oh spreading}," is not fully accurate. This inaccuracy stems from the initial transcription provided by Whisper, which included the word ``\textit{night}," leading our model to generate a sentence misaligned with the speaker's actual intent. This example underscores the importance of having accurate initial transcripts for improving the final output quality, especially in scenarios where subtle contextual nuances are critical.

It is important to note that the proposed model is capable of inferring and incorporating the speaker’s underlying intention more effectively. By leveraging gestures alongside incomplete or disfluent speech, our approach captures the full semantic meaning of the speaker’s communication. For example, when a speaker’s verbal output is fragmented or ambiguous, the model uses the accompanying gestures to infer the intended message, producing transcripts that are both richer and more contextually accurate. Unlike Whisper, which relies solely on the spoken words and often misses crucial gestural context, our method bridges the gap between speech and gesture, offering a more comprehensive representation of the speaker's intent. 

\section{Concluding Remarks}
In this paper, we proposed a novel approach utilizing multimodal LLMs to generate gesture-aware speech recognition transcripts for patients with language disorders. Our framework integrates verbal speech and iconic gestures, enabling the generation of enriched transcripts that capture the latent meaning conveyed through both modalities. Through extensive experimentation, we demonstrated that the proposed method effectively contextualizes incomplete or disfluent speech by incorporating gesture information, leading to more accurate and meaningful representations of the speaker's intent. These findings highlight the potential of our approach to significantly contribute to the field of speech and language therapy, offering innovative tools that can enhance the quality of life for individuals with language disorders by facilitating better communication and assessment methods.

\subsection{Ethical Statement} 
Our dataset was obtained from AphasiaBank with the approval of the Institutional Review Board (IRB) and adheres to the data sharing guidelines set by TalkBank\footnote{https://talkbank.org/share/ethics.html}. This includes complying with the Ground Rules for all TalkBank databases, which are based on the American Psychological Association Code of Ethics~\cite{american2002ethical}.

\subsection{Limitation \& Future Work} 
While the results are promising, we recognize several limitations and outline our plans to extend this work further.

One primary limitation is the absence of a definitive ground truth for quantitative evaluation. Since our model generates transcripts by synthesizing speech and gesture data from scratch, traditional benchmarks, such as comparisons with standard speech recognition outputs, are insufficient. Moreover, existing original transcripts lack gesture annotations, making direct comparisons challenging. In future work, we aim to address this gap by collaborating with certified pathologists to conduct qualitative assessments, such as A-B preference tests, to evaluate the effectiveness of gesture-enriched transcripts in accurately conveying the speaker's intentions.

To support quantitative evaluations, we plan to develop novel metrics that assess transcript quality, including grammar accuracy, semantic consistency, and the integration of multimodal information. Such metrics will provide a more objective basis for assessing our model's performance and facilitate comparisons with other multimodal and unimodal approaches.

Another limitation of this study is its focus on structured gestures from a specific task, the Peanut Butter Sandwich Task. While this task offers a controlled context for testing our approach, it does not encompass the diversity of gestures and communication patterns seen in everyday scenarios. As part of our future work, we plan to expand the scope of our model to include tasks such as the Cinderella Story Recall Task~\cite{bird1996cinderella}, which involves unstructured and complex narrative gestures. This expansion will allow us to evaluate the adaptability and robustness of our model in handling varied linguistic and gestural contexts.

In summary, while this study establishes a strong foundation for gesture-aware speech recognition, we aim to refine and extend our methods through collaborative qualitative evaluations, the development of robust quantitative metrics, and broader task applications. These efforts will ensure that our approach continues to evolve, ultimately contributing to more effective communication tools and interventions for individuals with language disorders.

\newpage
\section{Acknowledgements}
This research was supported by the MSIT (Ministry of Science and ICT), Korea, under the ICAN (ICT Challenge and Advanced Network of HRD) support program (IITP-2024-RS-2023-00259497) supervised by the IITP (Institute for Information \& Communications Technology Planning \& Evaluation) and under the Global Research Support Program in the Digital Field program (RS-2024-00425354) supervised by the IITP (Institute for Information \& Communications Technology Planning \& Evaluation).

\bibliography{reference}


\end{document}